\documentclass[useAMS,usenatbib]{mn2e}
\usepackage[]{txfonts}
\def\simless{\mathbin{\lower 3pt\hbox
{$\rlap{\raise 5pt\hbox{$\char'074$}}\mathchar"7218$}}}   
\def\simmore{\mathbin{\lower 3pt\hbox
{$\rlap{\raise 5pt\hbox{$\char'076$}}\mathchar"7218$}}}   
\newcommand{\be}{\begin{equation}}
\newcommand{\ee}{\end{equation}}
\usepackage{graphicx}
\usepackage{epsfig} 
\usepackage{epstopdf}

\title[S2: a probe of the accretion around Sgr A*]{The S2 star as a
probe of the accretion disk of Sgr A*}

\author[Dimitrios Giannios, Lorenzo Sironi]
{Dimitrios Giannios$^{1}$\thanks{E-mail: dgiannio@purdue.edu  (DG)}  and Lorenzo Sironi$^{2,3}$\thanks{E-mail: lsironi@cfa.harvard.edu (LS)}  \\
$^{1}$Department of Physics, Purdue University, 525 Northwestern
Avenue, West Lafayette, IN 47907, USA\\
$^{2}$Harvard-Smithsonian Center for Astrophysics, 60 Garden St., Cambridge, MA 02138, USA\\
$^{3}$NASA Einstein Postdoctoral Fellow}

\begin{document}
\date{Received / Accepted}
\pagerange{\pageref{firstpage}--\pageref{lastpage}} \pubyear{2013}

\maketitle

\label{firstpage}

\begin{abstract}
How accretion proceeds around the massive black hole in the Galactic
center and other highly sub-Eddington accretors remains poorly
understood. The orbit  of the S2 star  in the Galactic center passes through  the
accretion disk of the massive black hole and any observational
signature from such interaction may be used as an
accretion probe. Because of its early stellar type, S2 is expected to possess a fairly
powerful wind. We show here that the ram pressure of the accretion
disk shocks the stellar wind fairly close to the star. The shocked fluid reaches a temperature of 
$\sim 1\,$keV  and cools efficiently through optically thin,
thermal bremsstrahlung emission. The radiation from the shocked 
wind peaks around the epoch of the pericenter passage of the star at a luminosity 
potentially comparable to the quiescent emission detected from Sgr
A*. Detection of shocked wind radiation can constrain the density of
the accretion disk at a distance of several thousands of gravitational radii from the black hole.   
\end{abstract} 
  

\section{Introduction} 
\label{intro}

The compact radio source Sgr A* is believed to mark the location
of the massive black hole in the center of our Galaxy. The Galactic center is also 
observed as an IR and, possibly, X-ray source of luminosity
$\sim 10^{36}$erg s$^{-1}$ and $\simless \,3\times 10^{33}$erg s$^{-1}$,
respectively \citep{Genzel10}.
Both IR \citep{ghez04} and X-ray flaring \citep{baganoff01} is regularly observed on timescales ranging from minutes to hours. Flares are believed to be associated
with processes taking place close to the black-hole horizon. 

The radiation observed from Sgr A* is believed to be powered by the accretion
process. The radiative power is, however, a small fraction of the rate
at which gravitational energy is released by the gas accretion. The accretion is, therefore,
expected to take place through a hot, quasi-virialized, thick disk forming a
``Radiatively Inefficient Accretion Flow'' or RIAF. 
The fate of the non-radiated energy is model-dependent. The energy may
be advected into the black hole through an ``Advection-Dominated Accretion
Flow''  or ADAF \citep{Narayan95}; carried by convective motions in
a ``Convection-Dominated Accretion Flow'' or CDAF
\citep{Quataert00,Ball01}; or in kinetic form through winds
from the disk as in the ``Inflow-Outflow Solutions'' or ADIOS
\citep{BB99,BB04}. The various models make distinctly different
predictions for the gas density and its radial profile in the disk,
but convincing observational probes are still lacking. 
X-ray observations constrain the electron density close to the sphere of influence of the
black hole (i.e., at the Bondi radius $R_{\rm b}\simeq 0.04$ pc
$=2\times 10^5\,R_{\rm g}$, where $R_{\rm g}=GM_{\rm BH}/c^2$
is the gravitational radius for a black hole mass $M_{{\rm BH}}=4.3\times 10^{6}$\,M$_{\odot}$) to be $n_{\rm b}\sim
100$\,cm$^{-3}$ \citep{baganoff03}, but offer little clues for the gas properties  
within that radius, where the accretion disk is located. 

The region of $R<R_{\rm b}$ is not devoid of sources (besides Sgr
A*). It is filled with tens of massive stars, the so-called S cluster \citep{genzel03,gillessen09a}.
Most of these stars are B dwarfs and have elliptical orbits 
bringing them as close as $\sim$ a few$\,\times\, 10^3\,R_{\rm g}$ from
the black hole. B stars are also known to have powerful winds 
of substantial kinetic luminosities $L_{w}\simmore 10^{34}$ erg s$^{-1}$
and characteristic mass loss rates of $\sim 10^{-7}$\,M$_{\odot}$ yr$^{-1}$. 
Among the S-cluster stars, the S2 is characterized by both a close pericenter 
passage to Sgr A* and the earliest stellar type, possibly connected to
the most powerful wind in the cluster.  
As we show here, the interaction of the wind of S2 with the accretion disk
is expected to result in a characteristic rise of the X-ray emission
from Sgr A* around the epoch of pericenter passage on a timescale 
of months, that can be used to probe the gas properties at the Galactic center.

\section{The S2 star and the accretion disk in the Galactic center}

The S2 star is the brightest of the S cluster. It has a $\sim$16 year
orbit around Sgr A* \citep{gillessen09b} of eccentricity $e=0.88$ and a pericenter distance
of $R_{\rm p}=2800\,R_{\rm g}$. At pericenter, the velocity of the 
star reaches $v_{\rm p}\simeq \sqrt{2R_{\rm g}/R_{\rm p}}\,c\simeq
8\times 10^8$ cm s$^{-1}$.
The nature of the S2 star has been revealed in the study by
\citet{martins08}. S2 is an early B dwarf star (of spectral
type B0-2.5V) of intrinsic luminosity $\log L/L_{\odot}=4.2\div4.8$,
effective temperature $T=(20\div30)\,\times 10^3\,$K, and radius 
$R_* \simeq 10\,R_{\odot}$. The lack of particular wind absorption 
signatures in its spectrum places an upper limit on the wind mass 
loss rate of $\dot M\simless 3\times 10^{-7}v_8$\,M$_{\odot}$ yr$^{-1}$, 
where $v_w=10^{8}v_8$ cm s$^{-1}$ is the speed of the wind. 

The \citet{martins08} upper limit on the wind mass loss from S2 
is only modestly constraining. Early B stars
have, as a population, powerful and fast winds. Their wind velocity
ranges within $v_w=(0.5\div2)\times 10^8\,$cm s$^{-1}$, while  
their mass loss rate is in the range  
$\dot {M}= (0.3\div3)\times 10^{-7}$\,M$_{\odot}$ yr$^{-1}$ \citep{kudritzki00}. Here
we adopt $v_w=10^8\,$cm s$^{-1}$ and $\dot {M}=10^{-7}$\,M$_{\odot}$ yr$^{-1}$
as reference values for the wind from S2, with the understanding that $v_w$ is uncertain
by a factor of $\sim 2$ and $\dot {M}$ by a factor of $\sim 3$. 

The accretion rate of Sgr A* is quite uncertain and may well
be distance-dependent. Theoretical modeling and Faraday rotation measurements
 place limits in the range  $\dot{M}_{\rm BH}=3\times
 10^{-9}\div10^{-7}$\,M$_{\odot}$ yr$^{-1}$ for the accretion rate
at the black hole \citep{yuan03,marrone07,mosci09}. Given the estimated
accretion rate, a radiatively efficient disk would radiate
away $L\sim 0.1 \dot{M}_{\rm BH}c^2> 10^{37}$ erg s$^{-1}$. 
The bolometric luminosity from Sgr A* is $\sim 10^{36}$ erg s$^{-1}$
\citep{ghez04} suggesting that the most of the gravitational energy
released is likely not radiated away. Such an accretion flow is termed as RIAF. 
RIAFs are characterized by a thick $H/R\sim 1$ disk that is
partially pressure and partially rotation 
supported. The total pressure (sum of the ram and gas pressures) of 
the disk at distance $R$  from the black hole is
$P\sim \rho c^2 (R_{\rm g}/R)$, where $\rho (R)$ is the gas mass density.
This expression holds rather independently of the details of the
adopted model for the flow (ADAF, CDAF, or ADIOS). 
The electron density at the Bondi radius
$R_{\rm b}\sim 2\times 10^5R_{\rm g}$ is constrained directly by X-ray observations  to be $n_{\rm b}\sim 100$ cm$^{-3}$
(Baganoff et al. 2003). 
The gas density closer to the black hole is essentially unconstrained.
Possible scalings for the density profile adopted here are 
$n=n_{\rm b} (R_{\rm b}/R)^{3/2}$ as, e.g., motivated by the ADAF
solution \citep{Narayan95}; or
$n=n_{\rm b}  (R_{\rm b}/R)$ as, e.g., motivated by GRMHD
simulations \citep{mckinney12,sasha12,narayan12b}. In the case of a CDAF, the density profile is
 shallower $n\propto R^{-1/2}$ \citep{Quataert00}. At the pericenter
 of S2 (i.e., $R=R_{\rm p}\sim 0.01R_{\rm b}$), 
the density of the disk material is, therefore, expected to be 
$n\sim n_{\rm b} (R_{\rm b}/R_{\rm p})^{1/2\div3/2}\sim 10^3\div10^5$cm$^{-3}$.

\section{The stellar wind interaction with the accretion disk}

The interaction of the stellar wind with the accretion disk 
has two potential observational signatures. The stellar wind is terminated by a strong
shock  because of the confining pressure of the disk. The shocked gas
reaches a temperature of $\sim 1\,$keV  and cools via thermal bremsstrahlung 
emission mainly in the X-ray band. The RIAF also undergoes a
shock upon interaction with the stellar wind. The shock 
may accelerate non-thermal particles \citep{narayan12,sadowski13}. 
Inverse Compton scattering of the bright stellar photon
field by energetic electrons can power detectable
hard X-ray and $\gamma$-ray emission. 

\subsection{Thermal emission from the shocked wind}

While the stellar wind expands, its ram
pressure drops with distance as $P_w={\dot M} v_w/4\pi r^2$, 
where lower case $r$ is measured from the center of the star. The stellar wind is
terminated by a strong shock at a distance $r_{\rm sh}$ where the wind
ram pressure is balanced by that of the RIAF. The pressure of the 
RIAF is the sum of the thermal pressure and the ram pressure resulting
from the relative motion of the RIAF and the star. 
The thermal pressure is of order $P_{\rm th}\sim \rho c^2 (R_{\rm g}/2R)$
while the ram pressure is mainly a result of the stellar motion, so that $P_{\rm ram}\sim 2 \rho c^2 (R_{\rm g}/R)$.
The ram pressure varies by $\sim 50\%$ depending on the angle 
of the stellar and disk orbits. For the purpose of our estimates, we set the total disk
pressure as $P_{\rm tot}\sim 2.5 \rho c^2 (R_{\rm g}/R)$. Equating $P_{\rm tot}=P_w$
at $R=R_{\rm p}$ gives the distance at which the stellar wind is shocked, when S2 is at pericenter:
\be
r_{\rm sh}=6\times 10^{13}{\dot M}^{1/2}_{-7}v_8^{1/2}n_4^{-1/2}\quad \rm cm, 
\ee 
where $\dot{M}=10^{-7} {\dot M}_{-7} $\,M$_{\odot}$ yr$^{-1}$ is the stellar wind mass loss, and  $n=10^4n_4\,$cm$^{-3}$ is the electron (or proton) number density of the RIAF at pericenter.
Note that for reasonable parameters $r_{\rm sh}\ll R_{\rm p}$, so the
wind is terminated at a ``small'' distance from the S2 star.

In the post-shock region, the temperature of the gas is $T_{\rm sh}=1.3
\times 10^7v_8^2\,$K, and the electron number density is 
$n_{\rm sh}=4\,n_w(r_{\rm sh})=3\times 10^6\, n_4v_8^{-2}$cm$^{-3}$, where $n_w$ is the electron number density in the unshocked wind. Here, we have assumed solar metallicity (our best guess for the wind from a main
sequence star). The wind cools through optically thin, thermal bremsstrahlung emission.
For temperatures $T\sim 10^7$K and a solar-like composition, line emission dominates the cooling rate
over free-free emission by a modest factor. The  emissivity is approximately constant 
for $T\sim (0.3\div3)\times 10^7\,$K and of order 
$\Lambda_{\rm N}\sim 3\times 10^{-23}$ erg
cm$^3$s$^{-1}$\citep{Stevens92, Sutherland93}. The cooling
timescale of the plasma is then $t_{\rm c}= 2 k_{\rm B}T_{\rm sh}/n_{\rm
  sh}\Lambda_{\rm N} \simeq 4\times 10^7v_8^4n_4^{-1}$ s.
bremsstrahlung cooling competes with the adiabatic expansion 
of the shocked wind. The latter takes place on a timescale
 $t_{\rm exp}\sim r_{\rm sh}/v_{\rm sh}\sim 4r_{\rm sh}/v_{w}\simeq 2.4 \times 10^6 {\dot
  M}^{1/2}_{-7}v_8^{-1/2}n_4^{-1/2}$ s, where $v_{\rm sh}=v_w/4$ is the post-shock flow velocity.\footnote{The expansion time
of the shocked wind $t_{\rm exp}$  is somewhat shorter than the 
typical evolution time  $t_{\rm ev}=R_{\rm p}/v_{\rm p}$ of the stellar orbit at pericenter
(i.e., the time over which the confining disk pressure changes).  
We, therefore, assume that the wind-disk interface evolves
quasi-steadily along the stellar orbit.} Out of the total kinetic
luminosity $L_{w}={\dot M}v_w^2/2$ of the wind, 
a fraction $t_{\rm exp}/t_{\rm c}\simeq 0.06\,  {\dot
  M}^{1/2}_{-7}n_4^{1/2} v_8^{-9/2}$ is radiated in the X-ray band as
thermal bremsstrahlung. The X-ray luminosity of the shocked wind region
is
\be
L_X= \frac{t_{\rm exp}}{t_{\rm c}}L_{w}\sim 2\times 10^{33} 
{\dot M}^{3/2}_{-7}n_4^{1/2} v_8^{-5/2}\quad \rm erg~s^{-1}.
\label{eq:Lx}
\ee
Given that the quiescent X-ray luminosity from Sgr A* is at a similar
level \citep[a few $10^{33}$ erg/s;][]{baganoff03}, it is possible to measure
contributions to the total emission from the shocked wind 
when the S2 star is close to pericenter.

\begin{figure*}
\includegraphics[scale=0.25]{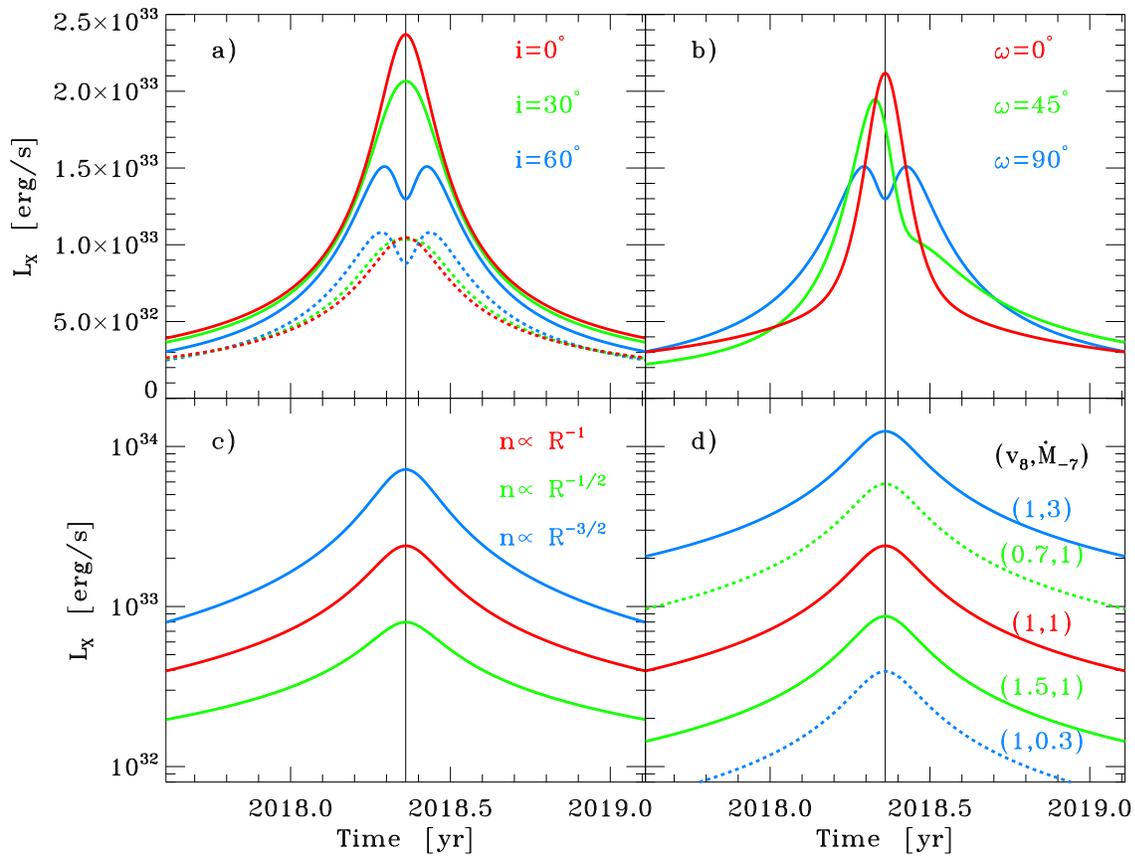}
\caption{Lightcurves of the X-ray emission from the shocked stellar
  wind, for different stellar and disk parameters.  In all the panels,
  the vertical black line marks the time of pericenter.~{\it (a)} We
  vary the inclination of the stellar orbit with respect to that of
  the accretion disk ($i=0^\circ$ in red, $i=30^\circ$ in green,
  $i=60^\circ$ in blue), for a fixed argument of periapsis
  ($\omega=90^\circ$). Solid lines show counter-rotating orbits,
  dashed for co-rotating orbits. ~{\it (b)} We vary the argument of
  periapsis of the stellar orbit ($\omega=0^\circ$ in red,
  $\omega=45^\circ$ in green, $\omega=90^\circ$ in blue), for a fixed
  orbital inclination ($i=60^\circ$). We only show counter-rotating
  orbits. ~{\it (c)} We vary the density profile in the accretion flow
  ($n\propto R^{-1}$ in red, $n\propto R^{-1/2}$ in green, $n\propto
  R^{-3/2}$ in blue), for a fixed density at the Bondi radius
  $n_b=130\, {\rm cm}^{-3}$. We only show counter-rotating orbits with
  $i=0^\circ$. ~{\it (d)} We vary the properties of the stellar wind
  with respect to our standard choice  $(v_{8},\dot{M}_{-7})=(1,1)$
  plotted in red, by changing the wind velocity (green)  or the mass
  loss rate (blue). We only show counter-rotating orbits with $i=0^\circ$.}
\label{fig:xray}
\end{figure*}

In Fig.~\ref{fig:xray}, we present the X-ray light curves expected
from the shocked wind of S2 when the star approaches its pericenter,
in mid-$2018$. For each point along the stellar orbit -- characterized
by the inclination $i$ and the argument of periapsis $\omega$ -- we
extract the local conditions in the disk from the model by
\citet[][see Table A1 there]{sadowski13}, which is based on GRMHD
simulations of the accretion flow around Sgr A*.\footnote{The
  inclination $i$ is such that $i=0^\circ$ if the orbital plane
  coincides with the disk equatorial plane. The argument of periapsis
  $\omega$ is measured along the orbital plane from the line of nodes,
  where the orbital plane intersects the disk midplane. We have
  $\omega=0^\circ$ if the pericenter is on the disk midplane.}  We
compute the shock radius $r_{\rm sh}$ at each time by balancing the
wind ram pressure with the total (ram plus thermal) pressure in the
accretion flow. When calculating the disk ram pressure as the star
moves through the disk, we differentiate between stellar orbits that
are co-rotating vs counter-rotating relative to the disk azimuthal
velocity. Given the shock radius $r_{\rm sh}$ at each time, we compute
the X-ray luminosity as outlined above.
 
In panels {\it (a)} and {\it (b)}, we show  how the X-ray emission
depends on the orientation of the stellar orbit relative to
the disk plane (we vary the inclination $i$ in panel {\it (a)}, for a
fixed argument of periapsis $\omega=90^\circ$; and we change the
argument of periapsis $\omega$ in panel {\it (b)}, for a fixed
inclination $i=60^\circ$). As shown in panel {\it (a)},
counter-rotating orbits (solid lines) always result in stronger X-ray
signatures than their co-rotating counterparts (dashed lines, with the
same color), since the ram pressure from the disk confines the shock
closer to the star, giving a higher post-shock particle density and
enhanced bremsstrahlung emission. The difference in the peak luminosity
is, however, modest (at most a factor of 2) when comparing counter and 
co-rotating orbits. The X-ray emission peaks when the
stellar orbit intersects the disk midplane, since the disk density is the
highest there (and so the shock is closest to the star). Depending on
the argument of periapsis, the star may cross the disk only once, at
pericenter (red line in  {\it (b)}, for $\omega=0^\circ$); or two
times, symmetric around the pericenter epoch (blue line in  {\it
  (b)}, for $\omega=90^\circ$); or two times, but asymmetric with
respect to pericenter (green line in  {\it (b)}, for
$\omega=45^\circ$). The shape of the light curve around pericenter can
then be used to constrain the  orientation of the orbit of S2 relative
to the accretion flow of Sgr A*.

The peak luminosity depends significantly on the structure of the
accretion flow around  Sgr A* and on the properties of the wind of
S2. In panel {\it (c)}, we fix the disk density at the Bondi radius
$n_b=130\,{\rm cm^{-3}}$ and we compare the scaling $n\propto R^{-1}$ expected
on the basis of GRMHD simulations \citep{mckinney12,sasha12,narayan12b,sadowski13} with the
prediction $n\propto R^{-3/2}$ of the ADAF solution (in blue) and with
the profile $n\propto R^{-1/2}$ of CDAF models (in green). In panel
{\it (d)}, we vary the velocity and mass loss rate of the stellar
wind, with respect to our reference choice $(v_{8},\dot{M}_{-7})=(1,1)$, plotted as
a  red curve. The trends seen in panels {\it (c)} and {\it
  (d)} can be simply understood from the scalings illustrated in
eq.~\ref{eq:Lx}, which gives the luminosity expected at pericenter. Once the properties of the stellar wind of S2 are
better determined, the X-ray emission from the shocked wind could be
used to place important constraints on the density profile of the
accretion flow around Sgr A*.

\subsection{Non-thermal radiation from the shocked
RIAF}

\citet{sadowski13} recently showed that a highly elliptical orbit of
an object moving through a RIAF is likely to be supersonic for the disk itself. 
The Mach number of the shock is $1.5\div3$ depending of the relative
orientation of the stellar and disk plane. The RIAF is shocked and 
electron acceleration has been proposed to take place at the shock
front \citep{sadowski13}. We explore this possibility by assuming 
that a fraction $\epsilon\sim 0.1$ of the total dissipated energy is deposited 
into non-thermal electrons with a rather flat power-law energy
distribution (i.e., an energy spectrum $dN/d\gamma\propto \gamma^{-2}$). Relativistic electrons will produce X-rays
and $\gamma$-rays by upscattering the powerful UV radiation 
from the S2 star. 

The RIAF material approaches the star with speed $v\sim v_{\rm p}$,
dissipating kinetic energy at a rate $L_{\rm diss}\sim \pi r_{\rm sh}^2
\rho v_{\rm p}^3\sim 10^{35}{\dot M}_{-7}v_8$erg s$^{-1}$.
The shocked RIAF material expands on a timescale $t_{\rm exp}\sim
4r_{\rm sh}/v_{\rm p}\simeq 3\times 10^5{\dot
  M}_{-7}^{1/2}v_8^{1/2}n_4^{-1/2}$s. A relativistic electron with
  Lorentz factor $\gamma$ cools on a timescale $t_{\rm c}=3\times
  10^7/\gamma U_{\rm ph}\,$s, where $U_{\rm ph}=L_*/4\pi r_{\rm sh}^2c$
is the energy density of the stellar radiation field, and we assume
that the S2 star has luminosity $L_*=10^{38}L_{*,38}$erg s$^{-1}$. The
characteristic energy of the stellar photons is $E_*\sim 3k_{\rm B}T_*\simeq 6\,
$eV, requiring electrons with Lorentz factor $\gamma=70\,E_{\rm 30keV}^{1/2}$ for the production of $\sim
30\,$keV photons. A fraction $t_{\rm exp}/t_{\rm
  c}\simeq 0.05 \,L_{*,38}E_{\rm 30keV}^{1/2}M_{-7}^{-1/2}v_8^{-1/2}n_4^{1/2}$ 
of the total dissipated power  $L_{\rm diss}$ is radiated away.
The luminosity of the IC component at photon energy $E$ is
\be
L_{\rm IC}\sim \epsilon  \frac{t_{\rm exp}}{t_{\rm c}}L_{\rm diss} 
\sim 6\times10^{32}\epsilon_{-1}L_{*,38}E_{30keV}^{1/2}
\dot{M}_{-7}^{1/2}v_8^{1/2}n_4^{1/2}\quad \rm erg~s^{-1}.
\ee
Currently the quiescent level of emission in the hard X-rays 
from Sgr A* is not known.  Extrapolation of the soft X-ray
emission spectrum indicates that the level of hard X-ray  luminosity at $\sim$30\,keV may be 
in the range $\sim 10^{32}\div3\times 10^{33}$erg/s. We conclude
that $L_{\rm IC}$ may be powerful enough to be detectable in this band 
over the quiescent emission from Sgr A*.  

Electrons with $\gamma\simmore\, 10^3M_{-7}^{1/2}v_8^{1/2} L_{38}^{-1}
n_4^{-1/2}$ are in the fast cooling regime ($t_{\rm exp}/t_{\rm
  cool}>1$).
As a result, all the energy that is injected in those electrons is radiated away
in the form of $E\simmore 6\,M_{-7}v_8L_{38}^{-2}
n_4^{-1}$MeV photons. Under the optimistic assumption of a 
flat injection electron distribution\footnote{If the electron
  power-law slope is $p>2$, the IC luminosity at a photon energy $E$
  will be suppressed by a factor of $\sim (E/E_*)^{(p-2)/2}$ with respect
  to the estimates given above.} up to $\gamma \simmore 10^4$, as
much as $L_{\rm IC}\sim 10^{34}\epsilon_{-1}{\dot M}_{-7}v_8$erg/s 
may be radiated in the MeV band [and up to
$\sim 200\,(6\,{\rm eV}/E_*)$\,GeV, where Klein-Nishina suppression appears].
 Unfortunately, the high density of sources in the Galactic center region makes
detecting such a  $\gamma$-ray signal very challenging.

\section{Discussion and Conclusions}
\label{conclusions}

Given its stellar type, S2 is expected to posses a moderately powerful stellar wind 
with $\dot {M}\sim 10^{-7}$M$_{\sun}$ yr$^{-1}$ and $v_w\sim 10^8$cm s$^{-1}$. 
Upon interaction with the accretion flow of Sgr A*, the stellar wind is shocked fairly close  to the
star, resulting in substantial bremsstrahlung cooling in the $\sim$a
few keV band. The predicted emission spectrum is 
that of optically thin, thermal bremsstrahlung, characterized by 
strong emission lines. 
The peak of the emission is predicted to take place around the 
pericenter passage of the star and the emission should remain bright 
over a several-month period (Fig.~1). The X-ray luminosity (eq.~2) is potentially
detectable close to its peak (or peaks) as an increament of the quiescent emission
that lasts for months\footnote{Current X-ray instruments do not have
 the angular resolution to distinguish this source from the Galactic
 center diffuse emission. Also the spectrum from the shocked wind
is similar (within observational uncertainty) to the quiescent
one from Sgr A* \citep{baganoff03}}. The slow evolution pattern of the
emission from the wind-disk interaction should allow
to discriminate it from the observed rapid (minute to hours) flaring that is
probably related to activity close to the black-hole horizon.
The shape of the light curve around the peak(s) depends on the
inclination or the orbit of S2 with respect to the disk and can
be used to determine the orbital plane of the disk. 

The wind-disk interaction leads to the formation of a shock in the disk material as well.
If the shock of the RIAF accelerates non-thermal electrons, then IC scattering of
the stellar radiation field leads to a significant hard X-ray signal.
NuSTAR \citep{H13} has the sensitivity to detect such a hard re-brightening   
from Sgr A* at the epoch of the pericenter passage of S2.

Similar considerations apply to the wind-disk interactions of
other S stars. From eq.~2, it is clear that high wind mass loss rates,
slow wind velocities and small pericenter radii
result in the brightest events. However, the other members of the
S cluster are dimmer than S2 and are likely to have weaker winds.  
For the moment S2 remains the best candidate for such a study. 

The next pericenter passage of S2  will take place in 2018. By that time,
a better observational understanding of the quiescent luminosity of Sgr A*
may be in place in both soft and hard X-rays. This will facilitate the
detection of even weak enhancements of the X-ray luminosity
of Sgr A* on a timescale of months to years 
around the closest approach of S2. Furthermore, our work motivates the need for a more precise
 determination of the wind properties of S2 (such as mass loss rate, velocity and 
metallicity). The current limit for the wind mass loss rate 
from S2 is ${\dot M}\simless 3\times 10^{-7}$M$_{\odot}$ yr$^{-1}$ \citep{martins08}, which is only modestly constraining. A precise assessment of the wind
properties will make any detection of shocked wind emission 
(or even upper limits) a very powerful probe of the accretion disk.  

In the Summer of 2013, the G2 cloud \citep{gillessen12,gillessen13} is passing at its pericenter 
in the disk of Sgr A* at a distance of $\sim 5000\,R_{\rm g}$. 
Interactions of the cloud with the accretion 
disk may lead to X-ray \citep{gillessen12} or radio \citep{narayan12,sadowski13} signatures that can
be used to probe the disk properties. An interesting fact 
is that the orbit of the S2 star is highly inclined with respect
to that of the G2 cloud, providing a potential probe
of the disk properties on a different plane.  

\section*{Acknowledgments}
L.S. is supported by NASA through Einstein
Postdoctoral Fellowship grant number PF1-120090 awarded by the Chandra
X-ray Center, which is operated by the Smithsonian Astrophysical
Observatory for NASA under contract NAS8-03060. 

\bibliography{S2probe.bib}

\end{document}